\documentstyle[psfig,twocolumn,prl,aps,amssymb]{revtex}
\begin{document}
\wideabs{
\draft

\title{Mixed States of Composite Fermions Carrying Two and Four Vortices}
\author{K. Park$^{\dagger}$ and J. K. Jain}
\address{Department of Physics, 104 Davey Laboratory,
The Pennsylvania State University,
Pennsylvania 16802}
\date{\today}

\maketitle

\begin{abstract}
There now exists preliminary experimental evidence for some fractions,
such as $\nu$ = 4/11 and 5/13, that do not belong to any of the
sequences $\nu=n/(2pn\pm 1)$, $p$ and $n$ being integers.
We propose that these states are mixed
states of composite fermions of different flavors, for example, composite
fermions carrying two and four vortices.
We also obtain an estimate of the lowest-excitation
dispersion curve as well as the transport gap; the gaps for 4/11
are smaller than those for 1/3 by approximately a factor of 50.
\end{abstract}

}


Two-dimensional electron systems exhibit spectacular phenomena
when subjected to an intense, perpendicular magnetic field. Most
remarkable is the
fractional quantum Hall effect (FQHE)~\cite{Tsui}, in which
the Hall resistance forms quantized plateaus at values
$R_H = h/ f e^2$ where $f$ is a simple rational fraction. The
prominent fractions appear according to the primary sequences,
\begin{equation}
f = \frac{n}{2pn \pm 1}
\end{equation}
where $p$ and $n$ are integers.  (The fractions $1-f$ are related
to these by particle-hole symmetry.)
An explanation of these sequences was one of the important initial 
successes
of the composite fermion (CF) theory; it
was in fact the clue that led to composite 
fermions~\cite{Jain,Review1,Review2}.
A composite fermion is the bound state of an electron
and an even number of quantum mechanical
vortices of the many body wave function
(sometimes thought of as an electron carrying an even number of
magnetic flux quanta, where a flux quantum is defined as $\phi_0=hc/e$).
The interacting electrons at Landau level (LL) filling factor
$\nu=\nu^*/(2p\nu^*\pm 1)$ transform into weakly
interacting composite fermions with vorticity $2p$ (denoted below by
$^{2p}$CFs) at an effective filling $\nu^*$.
The integral quantum Hall effect of composite fermions, corresponding
to $\nu^*=n$, manifests itself as the FQHE of electrons at
$f = \frac{n}{2pn \pm 1}$.
These states are ``pure", in the sense that they contain
only a single flavor of composite fermions, namely $^{2p}$CFs.

However, there now exist exceptions to the primary
states.  The FQHE at $\nu = 5/2$~\cite{Willett,Pan}
has been known for many years.  There is growing consensus that its
physical origin, while still formulated in terms of composite fermions,
is fundamentally distinct from the other, odd-denominator
fractions:  the 5/2 state is described in terms of a BCS-type
paired state of composite fermions~\cite{PairedCF},
arising because the residual interaction between composite
fermions is weakly attractive here~\cite{Cooper}, in contrast to
the other fractions that are described as
states containing an integral number of filled CF-LLs.
The focus of this article will be on $\nu =4/11$~\cite{nu4_11}. We
suggest that it has a more or less the traditional description in terms
of filled CF-LLs, except that it is a ``mixed" FQHE state of composite
fermions of two different flavors, those carrying two and four vortices.
(Here, the term ``mixed" refers to an admixture of two different CF
flavors, without necessarily implying spatial phase separation.)

Let us first see how the state at $\nu = 4/11$ is understood in
terms of a mixture of two different flavors of composite fermions.
Start by  considering the state of {\em fully} polarized electrons at 
$\nu=4/3$.
The state at $4/3=1+1/3$ is incompressible, at least for
certain class of interactions. It
contains one fully occupied Landau level of electrons
and the second Landau level at 1/3 filling.  The electrons in the second
Landau level are equivalent to composite fermions at effective filling
of unity.  Thus, the 4/3 state is the simplest, albeit somewhat trivial
example of a mixed state:
it contains one filled LL of $^0$CFs (composite fermions carrying
zero vortices, i.e., electrons) and one filled LL of $^2$CFs.  We
denote this state by $(\nu^{(0)},\nu^{(2)})=(1,1)$, where $\nu^{(2p)}$ is 
the
filling factor of $^{2p}$CFs.  Upon
attachment of two more vortices to each particle, a state at 4/11 is
obtained, which contains
both $^2$CFs and $^4$CFs, each at unit effective filling factor; in
other words, 4/11 is described as $(\nu^{(2)},\nu^{(4)})=(1,1)$.

W\'{o}js and Quinn~\cite{Quinn} searched for a fully polarized FQHE
at 4/11 numerically, through
exact diagonalization on an $N=8$ particle system.  They found no
gap in the excitation spectrum here; as a matter of fact,
the ground state here is not even uniform (it does not have
$L=0$ in the spherical geometry, where $L$ is the total angular momentum).
They concluded, based on this study, that there is no FQHE at 4/11,
at least for fully polarized electrons.
This result is not surprising in view of the fact that the fully
polarized state at 4/3 is rather fragile even for
electrons, quite close to an instability~\cite{highLLgap},
because the Coulomb matrix elements in the second Landau
level are less repulsive than those in the lowest
Landau level.  The attachment of two further vortices to each electron
to obtain the state at 4/11 would only further weaken it,
most likely destabilizing it altogether.

\begin{figure}
\centerline{\psfig{figure=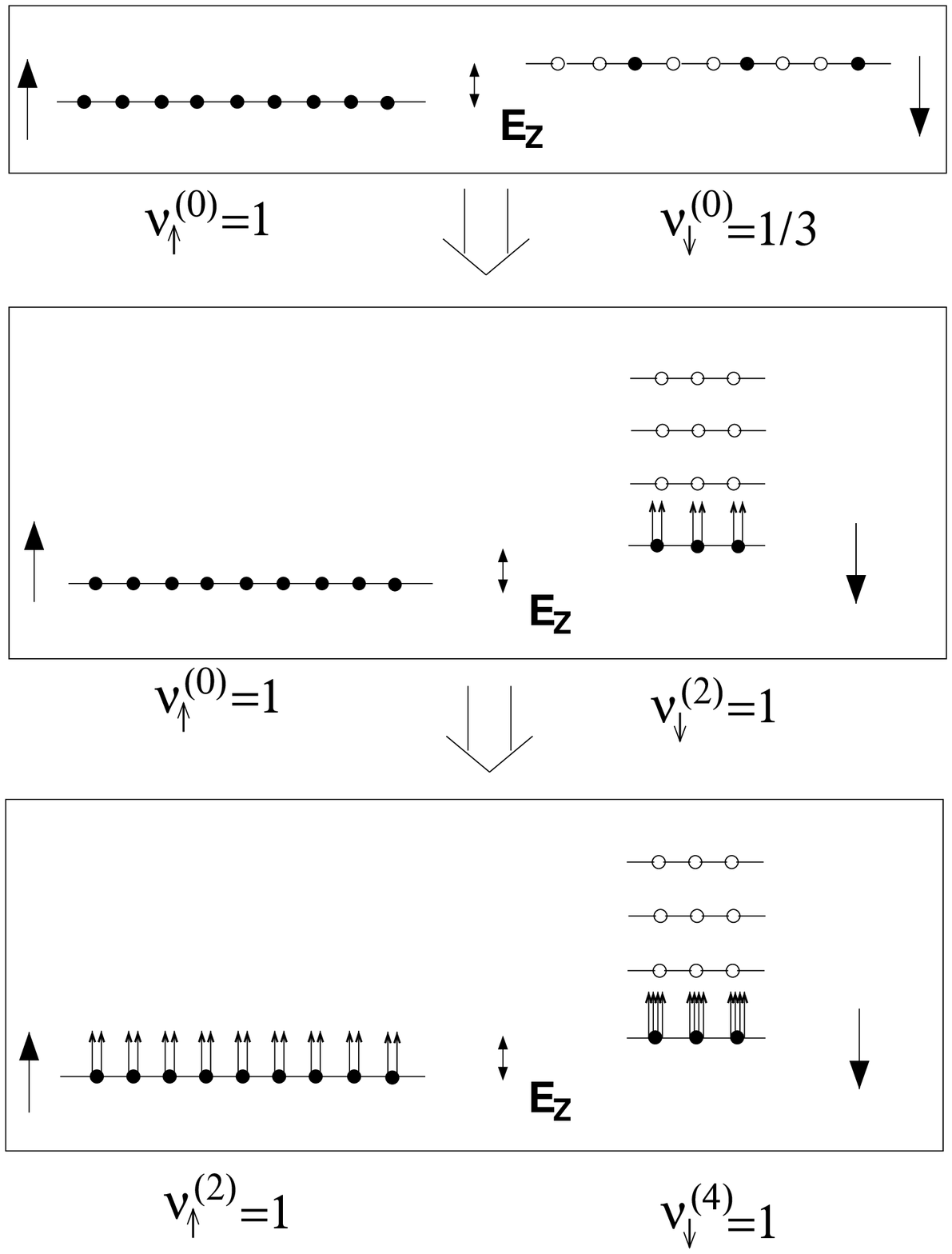,width=3in,angle=0}}
\vspace{0.2cm}
\caption{Schematic diagram explaining the physics of the
mixed CF state at $\nu = 4/11$.  Small arrows
decorating the circles depict the vortices captured by
composite fermions. Empty circles indicate empty
sites in a given Landau level. Big arrows near the Landau levels
signify the spin of the composite fermions.  The spin-up and
spin-down Landau levels are shifted in energy by the Zeeman splitting 
energy
$E_z$.  The top panel shows electrons at $\nu=4/3=1+1/3$
with the spin up Landau level fully occupied and the spin down Landau
level one third occupied.
The middle panel shows that the partially filled LL splits into
Landau levels of composite fermions, with 1/3 filling corresponding
to unit filling of $^2$CFs.  Attachment of two vortices to each
particle produces the partially polarized
4/11 state studied in this article (bottom panel),
which contains one filled $^4$CF-LL and one filled $^2$CF-LL, with two 
types of
composite fermions carrying opposite spins.  The filling factor of 
$^{2p}$CFs is
denoted by $\nu^{(2p)}$.
\label{diagram}}
\end{figure}

In order to resolve the apparent discrepancy between theory and experiment,
we consider a non-fully polarized FQHE state at 4/11.  At least
two such states are possible; our focus will be on the state
in which both spin up $^4$CFs and spin down $^2$CFs fill one
Landau level each: $(\nu_{\uparrow}^{(2)},\nu_{\downarrow}^{(4)})=(1,1)$.
(Here, the subscript of $\nu$ refers to the spin of the composite fermion.)
This state is related to $(\nu_{\uparrow}^{(0)},\nu_{\downarrow}^{(0)})
=(1,\frac{1}{3})$
as shown in Fig.~\ref{diagram}. The ground state
of this kind has been considered earlier
by MacDonald in the context of generalized
Laughlin states~\cite{MacDonald}.
It is the first member of the sequence
\begin{equation}
f'=\frac{1+f}{2(1+f)\pm 1}\;,
\end{equation}
with $f$ given in Eq.~(1).
For the following reasons, we believe that
$(\nu_{\uparrow}^{(2)},\nu_{\downarrow}^{(4)})=(1,1)$ will be
a stable FQHE state at 4/11 in an approprite range of Zeeman energies.
First, an exact
diagonalization study on sphere with $N=6$ electrons
tells us that the ground state at $\nu=4/11$ is an $L=0$ state
with partial polarization (to be specific,
total spin quantum number is $S = 1$)
even with a very small Zeeman splitting energy~\cite{Wu}.
Secondly, as we will see below, the wave functions of the composite fermion
theory obtain not only the correct spin and angular momentum
quantum numbers, but also accurate energies.
Finally, and most importantly, higher electronic
Landau levels are not used for the construction of this state, and
the argument given above regarding the instability of the
{\em fully polarized} 4/11 state is not effective here.  The partially
polarized state at 4/11 is expected to be more robust than the fully
polarized one for the same reason that the 1/3 state in
the second LL is rather weak but the 1/3 state in the
spin-reversed lowest LL is strong.

We will use the spherical geometry~\cite{Monopole} below, which
considers $N$ electrons on the surface of
a sphere in the presence of a radial
magnetic field emanating
from a magnetic monopole of strength $Q$, which corresponds to
a total flux of $2Q\phi_0$
through the surface of the sphere.
The wave function for the CF state at $Q$,
denoted by $\Psi_{2Q}$, is constructed by analogy to the wave function of
the corresponding electron states at $q$, denoted by $\Phi_{2q}$:
\begin{equation}
\Psi_{2Q} = {\cal P}_{LLL} \Phi_{N-1}^{2p} \Phi_{2q}
\end{equation}
Here $\Phi_{N-1}=\prod_{j<k}(u_j v_k - u_k v_j)$
is the wave function of the fully occupied
lowest Landau level with monopole strength equal to $(N-1)/2$,
where $u_j \equiv \cos(\theta_j /2)
\exp(-i\phi_j /2)$ and  $v_j \equiv \sin(\theta_j /2) \exp(i\phi_j /2)$.
${\cal P}_{LLL}$ denotes projection of the wave function into the lowest
Landau level (LLL).
The monopole strengths for $\Phi_{2q}$ and $\Psi_{2Q}$,
$q$ and $Q$, respectively, are related
by $Q=q+p(N-1)$. For the ground state and the single exciton
state, the wave functions $\Phi_{2q}$ are completely determined by
symmetry (i.e., by fixing the total orbital angular momentum $L$,
which is preserved in
going from $\Phi_{2q}$ to $\Psi_{2Q}$ according to the above rule),
giving parameter-free
wave functions $\Psi_{2Q}$
for the ground and single-exciton states of interacting electrons.
These have been found to be extremely accurate in tests against exact
diagonalization results available for small
systems~\cite{Review1,Review2,JK}.

To be concrete, we write a trial wave function
for the state at $\nu^* = 4/3$ as follows:
\begin{equation}
\Phi^{gr}_{\nu^*=4/3} = \prod_{i,j \in \uparrow}(u_iv_j-v_iu_j)
\prod_{k,l \in \downarrow}(u_kv_l-v_ku_l)^3
\label{eq:Phi4_3}
\end{equation}
where, for example,
$i \in \uparrow$ denotes that the $i$-th particle
is spin-up. Note that the spin part of the wave function
is not explicitly written; the full wave function is obtained by
multiplying the above wave function by the spin part and then
antisymmetrizing the product.
Upon the attachment of two vortices,
the CF wave function for the ground state
at $\nu=4/11$ is given by:
\begin{eqnarray}
\Psi^{gr}_{\nu=4/11} &=& \prod_{i,j \in \uparrow}(u_iv_j-v_iu_j)^3
\prod_{k,l \in \downarrow}(u_kv_l-v_ku_l)^5
\nonumber \\
&\times&
\prod_{m \in \uparrow, n \in \downarrow}(u_mv_n-v_mu_n)^2
\label{eq:Psi4_11gr}
\end{eqnarray}

Before proceeding further, let us make sure that
$\Psi^{gr}_{\nu=4/11}$ is an eigenstate of the total spin,
which can be shown as follows~\cite{MacDonald}.
First, $\Psi^{gr}_{\nu=4/11}$ has the same total-spin
eigenvalue as $\Phi^{gr}_{\nu^*=4/3}$
because  $\Psi^{gr}_{\nu=4/11}$
is obtained by multiplying $\Phi^{gr}_{\nu^*=4/3}$
by a symmetric polynomial.
Because the spin-up Landau level is full,
application of the total spin raising operator annihilates
$\Phi^{gr}_{\nu^*=4/3}$. Also, $\Phi^{gr}_{\nu^*=4/3}$
is evidently an eigenstate
of $S_z$, and therefore is an eigenstate of total spin
with $S = S_z = (N_\uparrow - N_\downarrow)\hbar/2$, where
$N_\uparrow$ and $N_\downarrow$ are the number of spin-up
and spin-down electrons, respectively.
This argument is valid for any state that has all single-particle orbitals 
of one
spin fully occupied.

\begin{figure}
\centerline{\psfig{figure=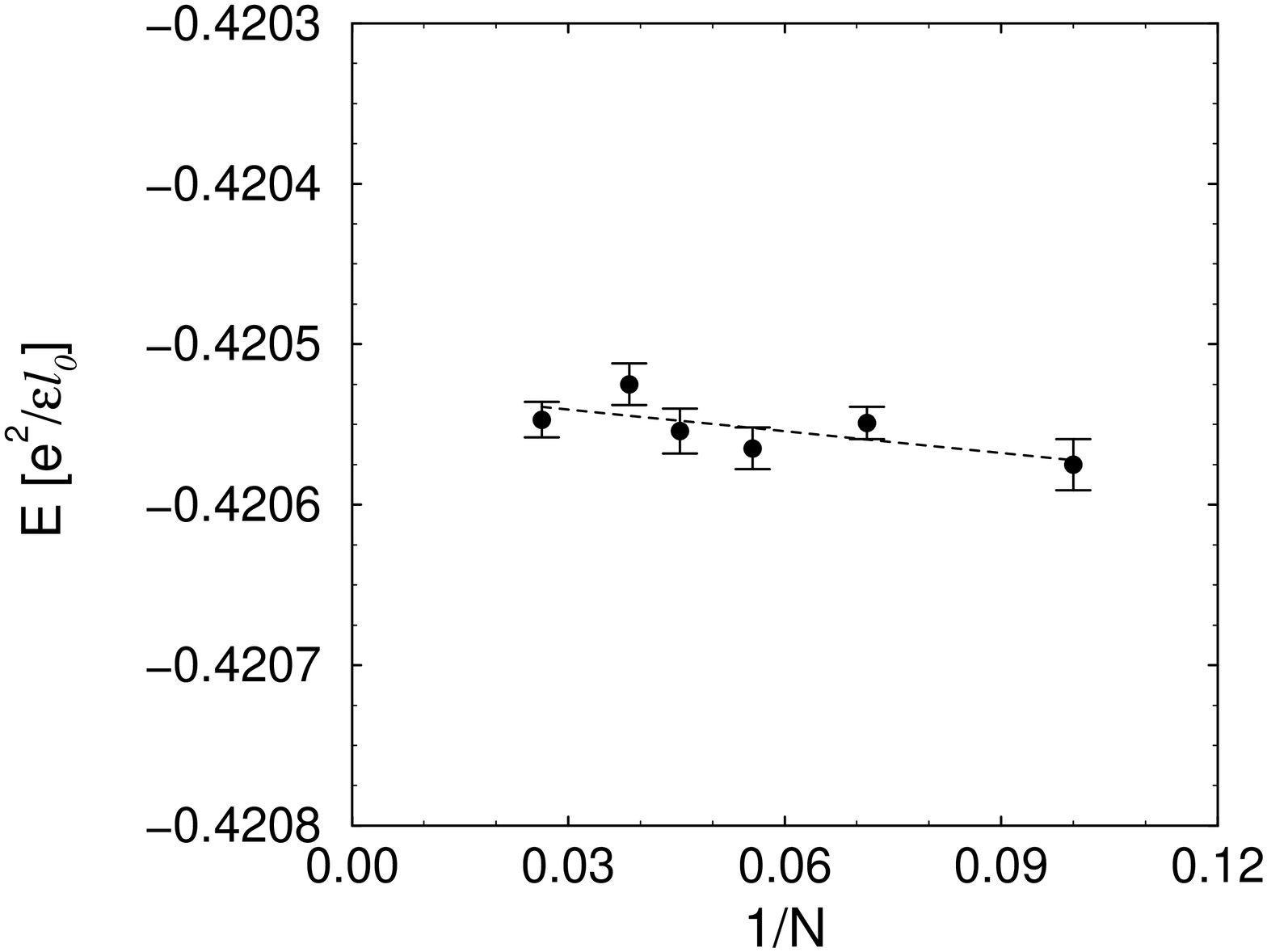,width=4.0in,angle=-90}}
\caption{Ground state energy at $\nu=4/11$ as a function of
$N^{-1}$, the number of electrons.  The quantity $l_0=\sqrt{\hbar c/eB}$
is the magnetic length, and
$\epsilon$ is the dielectric constant of the background material.
The error bars show one standard deviation in the Monte Carlo simulation.
\label{ground}}
\end{figure}

Having established that $\Psi^{gr}_{\nu=4/11}$ is a legitimate wave 
function,
we turn to the problem of energetics.
Fig.~\ref{ground} shows $N$-dependence of the energy
of the ground state wave function described by Eq.~(\ref{eq:Psi4_11gr}).
The pure Coulomb interaction $V(r)=e^2/\epsilon r$ is assumed here and 
below.
By using the linear extrapolation,
the ground state energy is estimated
to be $-0.420527(14)$ in units of $e^2/\epsilon l_0$
in the thermodynamic limit.  Here $l_0$ is the magnetic length at 
$\nu=4/11$ and
$\epsilon$ is the dielectric constant of the background material.
It is quite comparable to the energies of the fully polarized
states at 1/3 and 2/5 \cite {JK}.

In order to test the stability of this state, we consider its neutral and
charged excitations.  If it is found that an ``excitation" has lower energy
than the presumed ground state, we clearly have a wrong ``ground state".
While this procedure obviously cannot
capture every possible instability, it has proven to be extraordinarily 
powerful
in the past in ruling out FQHE states at low filling factors as well as in
higher Landau levels \cite{inst}.

The wave functions for the lowest-lying excitations are constructed
by promoting a $^4$CF into its lowest unoccupied $^4$CF-LL, while preserving
its spin.  Making an excitation in the $^2$CF part will produce a higher 
energy
excitation for the same reason as the excitation gaps are larger at
$n/(2n+1)$ than at $n/(4n+1)$.
Therefore, the wave function for excitations is
written as follows:
\begin{eqnarray}
\Psi^{ex}_{\nu=4/11}(L) &=&
\prod_{m \in \uparrow, n \in \downarrow}(u_mv_n-v_mu_n)^2>
\prod_{i,j \in \uparrow}(u_iv_j-v_iu_j)^3
\nonumber \\
&\times&
{\cal P}_{LLL}
\left[
\prod_{k,l \in \downarrow}(u_kv_l-v_ku_l)^4 \;
Det[\Phi^{ex}_{2q^*,\downarrow}(L)]
\right]
\label{eq:Psi4_11ex}
\end{eqnarray}
where $L$ is the total angular momentum and
$2 q^* = N_{\downarrow} - 1$. The number of spin-up
electrons is related to that of spin-down
electrons: $N_{\uparrow} = 3 N_{\downarrow} -2$. Of course,
if $\Phi^{ex}_{2q^*,\downarrow}$ is replaced by the ground-state
wave function at $q^*$,
$\Phi^{gr}_{2q^*,\downarrow} = \prod_{k,l\in\downarrow}(u_kv_l-v_ku_l)$,
Eq.~(\ref{eq:Psi4_11gr}) is obtained.
Comparison with exact diagonalization studies
sheds light on the accuracy of the above wave functions.
For $N=6$ system, the energies of the ground and excited
state are approximately 0.2 \% larger than
the exact energies; for $N=6$ and $Q=6.5$, the energies from
the wave functions are -0.473953(14) and -0.473611(16) $e^2/\epsilon l_0$ 
for
the ground and the excited states, respectively, which are to be compared
to the exact energies -0.4751 and -0.4742 $e^2/\epsilon l_0$~\cite{Wu}.

The energy gap of the lowest-lying excitations,
\begin{eqnarray}
\Delta(k) &=& \frac{\langle\Phi^{ex}_{\nu=4/11}(L)|V(r)
|\Phi^{ex}_{\nu=4/11}(L)\rangle}
{\langle\Phi^{ex}_{\nu=4/11}(L)|\Phi^{ex}_{\nu=4/11}(L)\rangle}
\nonumber \\
&-& \frac{\langle\Phi^{gr}_{\nu=4/11}|V(r)
|\Phi^{gr}_{\nu=4/11}\rangle}
{\langle\Phi^{gr}_{\nu=4/11}|\Phi^{gr}_{\nu=4/11}\rangle},
\end{eqnarray}
is computed using Monte Carlo methods in the
spherical geometry.
One of the most challenging aspects of the computation
stems from the fact that
the gap at $\nu=4/11$ is extremely small throughout the
whole dispersion of the excitation.  In fact,
it is the smallest gap ever calculated in the quantum Hall
effect; it is roughly 50  times
smaller than the gap at $\nu=1/3$.
As a result, the number of iterations of the Monte Carlo simulation
must be increased significantly
in order to minimize the statistical error, making the
computations tremendously more time-consuming than
for the primary states.  Typically, 100 million Monte Carlo iterations
were needed for each energy to obtain the desired accuracy, which
is an order or magnitude larger than the number of iterations used in
the studies of primary states ($\sim$ 10 million).
Another consequence of the smallness
of the gaps is that the intrinsic error in the gaps
is not negligible.  A comparison with exact diagonalization
studies (for 6 particles) shows that even though the energies
of the ground and excited state
are predicted correctly at the level of 0.2\%, the
gaps are reliable only to 10\%.
Such an error is acceptable in view of the significant
Monte Carlo uncertainty as well as our neglect of a number of
other effects that make much bigger corrections.

\begin{figure}
\centerline{\psfig{figure=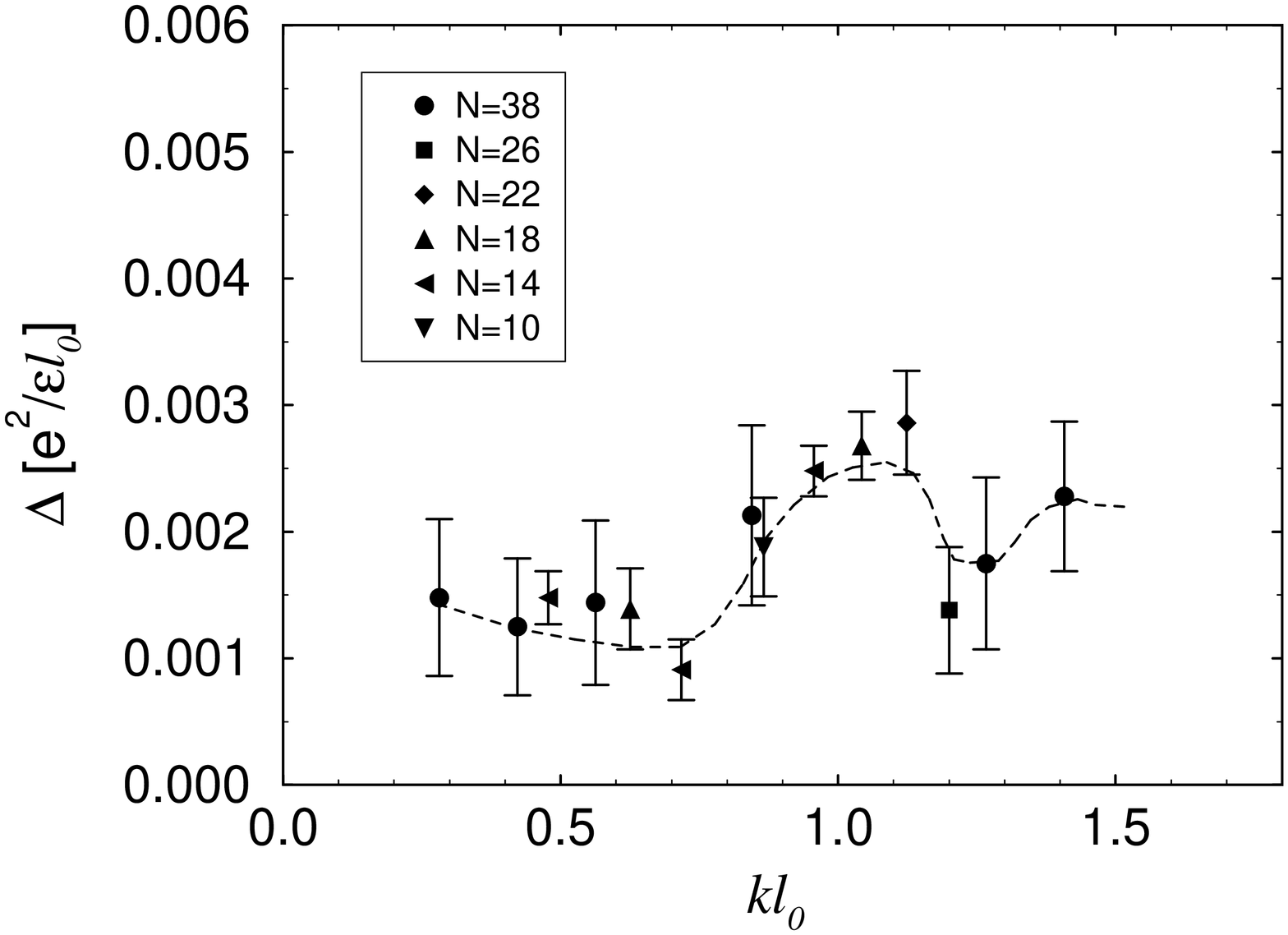,width=4.0in,angle=-90}}
\caption{Dispersion curve for the lowest excitations of
the partially polarized FQHE state
at $\nu=4/11$.  Several values of $N$ are used to determine the entire
curve. The dashed line is a guide to the eye.
\label{dispersion}}
\end{figure}

Fig.~\ref{dispersion} shows the dispersion curve
of the lowest excitation.
The results are plotted as a function of the
wave vector of the excitation, $k$, which is related to
the angular momentum $L$ via $k=L/R$ with $R$
being the radius of the sphere.  The transport gap,
which is the large wave vector limit of the dispersion curve,
is estimated to be 0.002(1) $e^2 / \epsilon l_0$.
Two roton minima are predicted in the dispersion
near $kl_0=$ 0.7 and 1.3 with energies of
around 0.001 $e^2/\epsilon l_0$.
While the full dispersion is in principle observable in
Raman scattering, the rotons may be easier to detect \cite{rotonexp}
The above numbers ought to be taken as
no more than rough estimates of the actual experimental gaps
because of the neglect in our calculation
of various realistic effects such as finite transverse
thickness, Landau level mixing and disorder. Previous
studies on the effects of finite thickness
and Landau level mixing~\cite{LLmix} give
30 to 50 \% reduction of the gap.  Therefore the
estimated gap at $\nu=4/11$ is even smaller than
the gap at $\nu =5/2$ \cite{Cooper}, whose Hall plateau
is firmly established only at ultra low
temperatures $\sim$ 4 mK~\cite{Pan}.

We end with a few comments.  First, a spin-{\em singlet} state at
4/11 can also be constructed, starting from the spin-singlet state
at 4/3 \cite{Wu2}.  It is likely that it has lower energy than the
one considered above at very small Zeeman energies, but it is
not possible to obtain reliable quantitative information for this
state due to the technical difficulties arising from the fact
that they involve inverse flux
attachment.  Second, a clear message of
our work is that the 4/11 FQHE is not fully polarized, which ought
to be testable in tilted field experiments.  Because the experiments
do observe a minimum at rather high magnetic fields, we suspect that
the actual observed state might be the partially spin polarized rather than
spin-singlet.  Finally, it is straightforward to enumerate other states
that will exhibit FQHE at the non-principal fractions; the ones
that are strongest are those that do not involve higher electronic
Landau levels in their construction (at intermediate steps -- of course,
all states are eventually projected into the lowest electronic Landau 
level).

>This work was supported in part by the National Science
Foundation under Grant No. DMR-9986806.
We thank V.W. Scarola for numerous helpful discussions,
W. Pan for sharing his data prior to publication,
and the Numerically Intensive Computing Group led by V.
Agarwala, J. Holmes, and J. Nucciarone,
at the Penn State University CAC for
assistance and computing time with the LION-X cluster.

$^{\dagger}$ Present address:
Department of Physics, Sloane Physics Laboratories, Yale
University, New Haven, CT 06520

\end{document}